\documentclass[aps,prd,preprint,superscriptaddress,nofootinbib]{revtex4-2}
\usepackage{graphicx}
\usepackage{bm}
\usepackage{amsmath}
\usepackage{amsfonts}
\usepackage{amssymb}
\usepackage[T1]{fontenc}
\usepackage[utf8]{inputenc}
\usepackage{newtxtext,newtxmath}
\usepackage{slashed}

\newcommand{\intk}{\int\frac{d^4k}{(2\pi)^4}}
\newcommand{\trD}{\mathrm{tr}_{D}}

\begin{document}

\title{Kaon gravitational form factors and mechanical structure in a three-flavor NJL model}

\author{Zhibo Liu}
\email{zhibo.liu.hep@gmail.com}
\affiliation{Department of Physics, Nagoya University, Nagoya 464-8602, Japan}

\author{Hiroaki Abuki}
\email{abuki@auecc.aichi-edu.ac.jp}
\affiliation{Department of Education, Aichi University of Education, Hirosawa, Kariya 448-8542, Japan}

\date{\today}

 \begin{abstract}
  
We investigate the gravitational form factors (GFFs) of the kaon in a
 proper-time-regularized three-flavor Nambu--Jona-Lasinio model.
We treat the kaon as a light--strange pseudoscalar bound state and evaluate
 its energy--momentum-tensor matrix element using a dressed quark--graviton vertex.
By retaining the interaction-induced contact contribution alongside the
 quark triangle diagrams, we preserve the gravitational Ward--Takahashi
 identity and ensure the conservation of the total energy--momentum tensor.
We determine the light- and strange-quark contributions to the kaon GFFs,
  contrast them with the corresponding pion
  results, and extract the associated pressure and shear-force distributions,
  together with the Breit-frame and transverse light-front mass radii.
Explicit flavor-symmetry breaking dictates that the strange sector
 carries a larger share of the kaon momentum and mechanical response.
Consequently, compared to the pion, the kaon exhibits a less negative
 $D$-term and more compact mechanical distributions, featuring enhanced
 central pressure and shear.
We further show that a Ward--Takahashi-identity-preserving reduction
 of the $C$-term projection is essential for maintaining the pion
 low-energy theorem in proper-time regularization.
 \end{abstract}

\maketitle

\section{Introduction}\label{sec:introduction}
  
Matrix elements of the QCD energy--momentum tensor (EMT) provide a
description of hadron structure complementary to that obtained from
electromagnetic and weak currents.
The Lorentz decomposition of these matrix elements defines the gravitational
form factors (GFFs), which encode how energy, momentum, angular momentum,
and internal forces are distributed inside a hadron~\cite{Kobzarev:1962wt,%
Pagels:1966zza,Ji:1996ek,Polyakov:2002yz,Polyakov:2018zvc}.
For a spin-zero target, the conserved EMT is parametrized by two independent
form factors.
The momentum form factor $A(t)$ obeys $A(0)=1$, whereas the $D$-term form
factor, denoted by $C(t)$ in this work, governs the spatial stress tensor
and hence the pressure and shear-force distributions~\cite{Polyakov:1999gs,%
Teryaev:1999su,Goeke:2007fp,Cebulla:2007ei,Lorce:2018egm,Hudson:2017oul}.

GFFs are related to Mellin moments of generalized parton distributions
(GPDs)~\cite{Mueller:1994ses,Ji:1996nm,Radyushkin:1997ki,Collins:1996fb,
Diehl:2003ny,Belitsky:2005qn}.
This connection has motivated experimental extractions, lattice QCD
calculations, and a broad range of continuum studies~\cite{Burkert:2018bqq,%
Burkert:2023wzr,Dutrieux:2021nlz,Hagler:2007xi,Pefkou:2021fni,Hackett:2023nkr,%
Hackett:2023rif}.
Pseudoscalar mesons are particularly instructive because their GFFs are
constrained by chiral symmetry and can be studied in approaches that preserve
the relevant Ward identities.
The pion has therefore served as the primary benchmark for analyses based
on chiral quark models, Dyson--Schwinger equations, light-front methods, and
related continuum frameworks~\cite{Broniowski:2008hx,Freese:2019bhb,%
Raya:2021zrz,Xu:2023ext,Sultan:2024cba}.

The kaon adds a qualitatively new element: explicit $SU(3)_F$ symmetry breaking.
Its unequal light- and strange-quark masses make it the simplest pseudoscalar
system in which one can resolve how flavor asymmetry redistributes momentum
and internal forces. 
Although recent continuum, dispersive, and light-front studies
have begun to quantify this effect~\cite{Liu:2025kaon,Xu:2023ext,
Cao:2025xhk,Sain:2026blfq}, a symmetry-preserving low-energy calculation
remains highly valuable for isolating the mechanisms that distinguish the
kaon from the pion and for identifying which features are robust against
the choice of spatial representation.

The Nambu--Jona-Lasinio (NJL) model offers an ideal and transparent setting
for this purpose, as it elegantly implements dynamical chiral symmetry breaking
while maintaining tractability.
It has been used extensively to describe meson spectra, decay constants, form
factors, and parton distributions~\cite{Nambu:1961tp,Nambu:1961fr,Vogl:1991qt,%
Klevansky:1992qe,Hatsuda:1994pi,Buballa:2003qv,Bentz:2001vc,%
Cloet:2005rt,Bentz:2009yy,Cloet:2014rja}.
However, evaluating the EMT within a contact interaction requires more than
merely attaching the EMT to the propagating quark lines.
The quark--graviton vertex must satisfy the gravitational Ward--Takahashi
identity (GWTI), and the local four-fermion interaction produces an additional
contact diagram.
Omitting either contribution leads to a nonconserved total EMT and can leave
regulator-dependent artifacts in the $D$-term channel.

In this work, we calculate the kaon GFFs in a proper-time-regularized
three-flavor NJL model without the Kobayashi--Maskawa--'t Hooft determinant
interaction, using the parameter set from Ref.~\cite{CarrilloSerrano:2014}.
We construct the dressed quark--graviton vertex, retain the
interaction-induced contact contribution, and implement the $C$-term
projection only after a GWTI-preserving reduction of the loop integrand.
The quark GWTI, together with canonical Bethe--Salpeter normalization, enforces
$A_K(0)=1$. The triangle and contact contributions then combine, through the
kaon pole condition, to cancel the nonconserved form factor $\bar c_K(t)$.
In the $C$-term channel, carrying out the GWTI reduction before imposing the
proper-time cutoff removes a local regulator artifact and rigorously recovers
the soft-pion low-energy theorem ($C_\pi(0)=-1$).
We then separate the light- and strange-quark contributions, compare the
kaon with the pion, and analyze the resulting mechanical distributions
and mass radii.

The paper is organized as follows.
Section~\ref{sec:njl} summarizes the three-flavor NJL framework and the kaon
bound-state amplitude.
Section~\ref{sec:quark_gffs} constructs the dressed quark--graviton vertex.
Section~\ref{sec:kaon_gffs} presents the conserved kaon EMT, the form factors,
and the associated mechanical structure.
Finally, Section~\ref{sec:summary} collects the main conclusions and discusses
the limitations of this work.

\section{NJL framework and kaon bound state}\label{sec:njl}
We use a three-flavor NJL model in the isospin-symmetric limit.
The interaction channels relevant to the present calculation are contained in
\begin{align}
\mathcal L_{\rm NJL}
&=
\bar\psi\left(i\slashed{\partial}-\hat m\right)\psi
+
\frac{G_\pi}{2}\sum_{a=0}^{8}
\left[
    \left(\bar\psi\lambda_a\psi\right)^2
    +
    \left(\bar\psi i\gamma_5\lambda_a\psi\right)^2
\right]
\nonumber\\
&\quad
-
\frac{G_\rho}{2}\sum_{a=1}^{8}
\left[
    \left(\bar\psi\gamma^\mu\lambda_a\psi\right)^2
    +
    \left(\bar\psi\gamma^\mu\gamma_5\lambda_a\psi\right)^2
\right]
-
\frac{G_f}{2}
\left(\bar\psi\gamma^\mu\gamma_5\psi\right)^2.
\label{eq:NJL_Lagrangian}
\end{align}
Here $\psi=(u,d,s)^T$ and $\hat m=\mathrm{diag}(m_\ell,m_\ell,m_s)$,
with $m_\ell\equiv m_u=m_d$.
The flavor matrices satisfy $\mathrm{tr}(\lambda_a\lambda_b)=2\delta_{ab}$
and $\lambda_0=\sqrt{2/3}\,\mathbf 1$.
We do not include the determinant interaction.
Consequently, the gap equation remains diagonal in flavor space, and explicit flavor
breaking enters through $m_\ell\neq m_s$ and the resulting constituent masses.

Since the model is nonrenormalizable, we define loop integrals using proper-time
regularization.
After Wick rotation, we use
\begin{equation}
    \frac{1}{X^n}
    \longrightarrow
    \frac{1}{(n-1)!}
    \int_{1/\Lambda_{\rm UV}^2}^{1/\Lambda_{\rm IR}^2}
    d\tau\,\tau^{n-1}e^{-\tau X}.
\label{eq:proper_time_regularization}
\end{equation}
The ultraviolet cutoff fixes the short-distance scale of the effective
model, whereas the infrared cutoff removes unphysical thresholds for the
decay of color-singlet bound states into constituent quarks.

At mean-field level, the dressed propagator is
\begin{equation}
    S_f(k)
    =
    \frac{1}{\slashed{k}-M_f+i\epsilon}
    =
    \frac{\slashed{k}+M_f}{k^2-M_f^2+i\epsilon},
    \qquad f=\ell,s,
\label{eq:quark_propagator}
\end{equation}
and the constituent mass satisfies
\begin{equation}
    M_f
    =
    m_f
    +
    \frac{3G_\pi M_f}{\pi^2}
    \int_{1/\Lambda_{\rm UV}^2}^{1/\Lambda_{\rm IR}^2}
    d\tau\,\frac{e^{-\tau M_f^2}}{\tau^2}.
\label{eq:gap_equation}
\end{equation}

The kaon appears as a pole in the light--strange pseudoscalar scattering amplitude.
Defining
\begin{equation}
    \Pi_{PP}^{fg}(q^2)
    =
    i(2N_c)
    \int\frac{d^4k}{(2\pi)^4}
    \trD\!
    \left[
        \gamma_5S_f(k)\gamma_5S_g(k+q)
    \right],
\label{eq:PP_polarization}
\end{equation}
the resummed amplitude is
\begin{equation}
    T_{PP}^{fg}(q)
    =
    \frac{-2iG_\pi}{1+2G_\pi\Pi_{PP}^{fg}(q^2)}.
\label{eq:PP_T_matrix}
\end{equation}
The kaon mass and pole residue are therefore fixed by
\begin{align}
    1+2G_\pi\Pi_{PP}^{\ell s}(m_K^2)&=0,
\label{eq:PP_pole_condition}
\\
    Z_K^{-1}
    &=
    -\left.
    \frac{\partial\Pi_{PP}^{\ell s}(q^2)}{\partial q^2}
    \right|_{q^2=m_K^2},
    \qquad g_{Kqq}=\sqrt{Z_K}.
\label{eq:Z_K}
\end{align}

Furthermore, two additional quark bubbles enter the dressed graviton
vertex, which we define as
\begin{align}
    \Pi_{SS}^{ff}(q^2)
    &=
    i(2N_c)
    \int\frac{d^4k}{(2\pi)^4}
    \trD\!
    \left[S_f(k)S_f(k+q)\right],
\label{eq:SS_polarization}
\\
    \Pi_{AA}^{ff,\mu\nu}(q)
    &=
    i(2N_c)
    \int\frac{d^4k}{(2\pi)^4}
    \trD\!
    \left[
        \gamma^\mu\gamma_5S_f(k)
        \gamma^\nu\gamma_5S_f(k+q)
    \right].
\label{eq:AA_polarization_tensor}
\end{align}
The transverse coefficient of the latter is denoted by
$\Pi_{AA,T}^{ff}(q^2)$.

The parameter set is listed in Table~\ref{tab:NJL_parameters}.
The vector coupling is part of the underlying parameterization,
although it does not enter the kaon GFFs explicitly at the order
considered here.

\begin{table}[t]
\caption{Model parameters used in the calculation, taken from
Ref.~\cite{CarrilloSerrano:2014}.
Mass scales are given in GeV and four-fermion couplings in ${\rm GeV}^{-2}$.}
\label{tab:NJL_parameters}
\begin{ruledtabular}
\begin{tabular}{ccccccc}
$\Lambda_{\rm IR}$
&
$\Lambda_{\rm UV}$
&
$M_\ell$
&
$M_s$
&
$G_\pi$
&
$G_\rho$
&
$G_f$
\\
\hline
0.240 & 0.645 & 0.40 & 0.59 & 19.0 & 10.8 & 0.82
\end{tabular}
\end{ruledtabular}
\end{table}

With the parameters in Table~\ref{tab:NJL_parameters}, the kaon pole
condition yields $m_K=0.469~{\rm GeV}$ and Eq.~\eqref{eq:Z_K} determines
$Z_K=20.5$ in the bubble normalization of Eq.~\eqref{eq:PP_polarization}.
For the pion, we use $m_\pi=0.138~{\rm GeV}$ and evaluate the same
residue definition in the equal-mass light-quark channel at this mass,
which results in $Z_\pi=17.9$.

\section{Dressed quark gravitational form factors}\label{sec:quark_gffs}
We next couple the canonical EMT to a dressed quark.
For incoming and outgoing quark momenta $p$ and $p'$, respectively,
the dressed quark--graviton vertex $\Gamma_{G,f}^{\mu\nu}(p',p)$ satisfies
\begin{equation}
    \Delta_\mu\Gamma_{G,f}^{\mu\nu}(p',p)
    =
    p^\nu S_f^{-1}(p')-p'^\nu S_f^{-1}(p),
    \qquad \Delta=p'-p.
\label{eq:GWTI_quark}
\end{equation}
In a local four-fermion theory, the inhomogeneous Bethe--Salpeter equation
for this vertex is algebraic.
Its driving term must include the five-point coupling generated by the
interaction; this term is required to satisfy Eq.~\eqref{eq:GWTI_quark}.

Writing $k=(p+p')/2$ and $t=\Delta^2$, the solution can be parameterized as
\begin{align}
    \Gamma_{G,f}^{\mu\nu}
    \left(k+\frac{\Delta}{2},k-\frac{\Delta}{2}\right)
    &=
    \gamma^\mu k^\nu
    -
    g^{\mu\nu}(\slashed{k}-M_f)
\nonumber\\
    &\quad
    +
    \frac{\Delta^\mu\Delta^\nu-\Delta^2g^{\mu\nu}}{4M_f}
    C_f^Q(t)
    +
    \frac{i\epsilon^{\mu\nu\alpha\beta}
    \Delta_\alpha\gamma_\beta\gamma_5}{4}
    D_f^{0,Q}(t).
\label{eq:dressed_quark_graviton_vertex}
\end{align}
The two interaction-generated form factors are transverse to $\Delta_\mu$
and are given by
\begin{equation}
    C_f^Q(t)
    =
    -\frac{8G_\pi\Pi_{SG}^{f}(t)}
    {1-2G_\pi\Pi_{SS}^{ff}(t)},
    \qquad
    D_f^{0,Q}(t)
    =
    \frac{2G_f\Pi_{AA,T}^{ff}(t)}
    {1+2G_f\Pi_{AA,T}^{ff}(t)}.
\label{eq:quark_vertex_form_factors}
\end{equation}
Here the scalar--graviton bubble is defined through
\begin{align}
    \frac{tg^{\mu\nu}-\Delta^\mu\Delta^\nu}{M_f}
    \Pi_{SG}^{f}(t)
    &=
    iN_c\intk\trD
    \left[
        S_f\left(k+\frac{\Delta}{2}\right)
        \Gamma_{G,f}^{\mu\nu,0}
        S_f\left(k-\frac{\Delta}{2}\right)
    \right],
\label{eq:Pi_SG_definition}
\\
    \Gamma_{G,f}^{\mu\nu,0}
    &=
    \gamma^\mu k^\nu-g^{\mu\nu}(\slashed{k}-M_f).
\nonumber
\end{align}

For later comparison, the canonical EMT matrix element of an on-shell quark
is decomposed as
\begin{align}
    \langle q_f(p')|T_f^{\mu\nu}(0)|q_f(p)\rangle
    &=
    \bar u_f(p')
    \Bigg[
        \frac{k^\mu k^\nu}{M_f}A_f^Q(t)
        +
        \frac{ik^{\{\mu}\sigma^{\nu\}\rho}\Delta_\rho}{2M_f}
        \left(A_f^Q(t)+B_f^Q(t)\right)
\nonumber\\
    &\qquad
        +
        \frac{\Delta^\mu\Delta^\nu-\Delta^2g^{\mu\nu}}{M_f}
        C_f^Q(t)
        +
        M_fg^{\mu\nu}\bar c_f^Q(t)
        +
        \frac{ik^{[\mu}\sigma^{\nu]\rho}\Delta_\rho}{2M_f}
        D_f^Q(t)
    \Bigg]u_f(p),
\label{eq:quark_EMT_decomposition}
\end{align}
where $k=(p+p')/2$ and braces (brackets) denote symmetrization
(antisymmetrization) with a factor of $1/2$.
Equation~\eqref{eq:dressed_quark_graviton_vertex} then yields
\begin{equation}
    A_f^Q(t)=1,
    \qquad
    B_f^Q(t)=0,
    \qquad
    \bar c_f^Q(t)=0,
    \qquad
    D_f^Q(t)=-1+D_f^{0,Q}(t).
\label{eq:quark_GFF_summary}
\end{equation}

\section{Kaon gravitational form factors and mechanical structure}\label{sec:kaon_gffs}
\subsection{Conserved kaon energy--momentum tensor}
For a spin-zero kaon, the EMT matrix element is written as
\begin{align}
    \langle K(p')|T^{\mu\nu}(0)|K(p)\rangle
    &=
    2P^\mu P^\nu A_K(t)
    +
    \frac{1}{2}
    \left(\Delta^\mu\Delta^\nu-\Delta^2g^{\mu\nu}\right)C_K(t)
    +
    2m_K^2\bar c_K(t)g^{\mu\nu},
\label{eq:kaon_EMT_decomposition}
\end{align}
where $P=(p+p')/2$ and $t=\Delta^2$.
In our convention, $C_K(t)$ is the kaon $D$-term form factor.
Conservation of the total EMT requires $\bar c_K(t)=0$, while the momentum
sum rule gives $A_K(0)=1$.

\begin{figure}[t]
 \centering
 \includegraphics[width=0.98\linewidth]{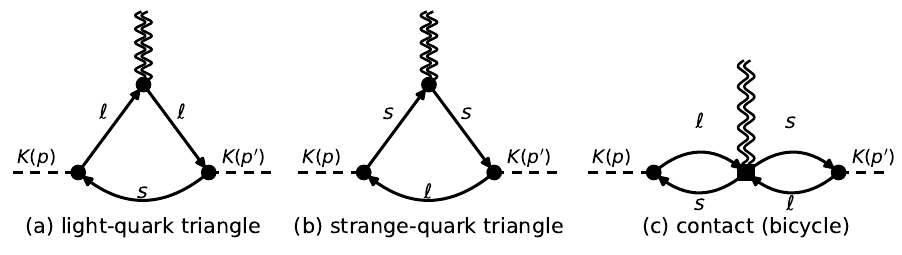}
 \caption{Feynman diagrams contributing to the kaon energy--momentum tensor 
 matrix element: (a) the EMT insertion on the light-quark line,
 (b) the EMT insertion on the strange-quark line, and (c) the
 interaction-induced contact (bicycle) diagram.
 Solid lines denote dressed quarks, dashed lines the external kaons,
 and the double wavy line the EMT insertion.}
 \label{fig:diagrams}
\end{figure}

The kaon Bethe--Salpeter vertex is
\begin{equation}
    \Gamma_K=\gamma_5g_{Kqq}\lambda_K,
    \qquad g_{Kqq}=\sqrt{Z_K}.
\label{eq:kaon_BS_vertex}
\end{equation}
We use
$\lambda_K=(\lambda_4+i\lambda_5)/\sqrt{2}$, so that
$\mathrm{tr}(\lambda_K\lambda_K^\dagger)=2$.
With $k_\pm=k\pm\Delta/2$, the two triangle contributions are
\begin{align}
    \mathcal M_{\ell,\triangle}^{\mu\nu}
    &=
    i(2N_c)Z_K\intk\trD
    \left[
        \gamma_5S_s(k-P)\gamma_5S_\ell(k_+)
        \Gamma_{G,\ell}^{\mu\nu}(k_+,k_-)S_\ell(k_-)
    \right],
\label{eq:kaon_light_triangle}
\\
    \mathcal M_{s,\triangle}^{\mu\nu}
    &=
    i(2N_c)Z_K\intk\trD
    \left[
        \gamma_5S_\ell(k-P)\gamma_5S_s(k_+)
        \Gamma_{G,s}^{\mu\nu}(k_+,k_-)S_s(k_-)
    \right].
\label{eq:kaon_strange_triangle}
\end{align}
In addition to the triangle diagrams, coupling the energy--momentum tensor
directly to the local four-fermion interaction generates a contact, or
bicycle, contribution $\mathcal M_{\rm B}^{\mu\nu}$.
The corresponding five-point vertex in the kaon pseudoscalar channel is
\begin{equation}
    \gamma_{G,PP}^{\mu\nu}
    =-2G_\pi g^{\mu\nu}
    \left(i\gamma_5\lambda_K\right)
    \otimes
    \left(i\gamma_5\lambda_K^\dagger\right).
\label{eq:kaon_five_point_vertex}
\end{equation}
It joins two factorized pseudoscalar bubbles and gives
\begin{align}
    \mathcal M_{\rm B}^{\mu\nu}(p',p)
    &=
    2G_\pi Z_K g^{\mu\nu}
    \Pi_{PP}^{\ell s}(p'^2)\Pi_{PP}^{s\ell}(p^2)
\nonumber\\
    &\xrightarrow{\ p^2=p'^2=m_K^2\ }
    2G_\pi Z_K g^{\mu\nu}
    \left[\Pi_{PP}^{\ell s}(m_K^2)\right]^2.
\label{eq:kaon_bicycle}
\end{align}
These three distinct topologies---the light-quark triangle, the strange-quark
triangle, and the interaction-induced contact diagram---are illustrated
in Fig.~\ref{fig:diagrams}.
Summing these pieces, the full matrix element reads
\begin{equation}
    \mathcal M_K^{\mu\nu}
    =
    \mathcal M_{\ell,\triangle}^{\mu\nu}
    +
    \mathcal M_{s,\triangle}^{\mu\nu}
    +
    \mathcal M_{\rm B}^{\mu\nu}.
\label{eq:kaon_total_matrix_element}
\end{equation}

Applying the quark GWTI before regularization, the sum of the two triangle
contributions to the coefficient of $g^{\mu\nu}$ is $Z_K\Pi_{PP}^{\ell s}(m_K^2)$.
For unequal constituent masses, the two triangle contributions need not be
equal separately.
The bicycle contributes $2G_\pi Z_K[\Pi_{PP}^{\ell s}(m_K^2)]^2$, and the
nonconserved part of the complete matrix element therefore reduces to
\begin{equation}
    2m_K^2\bar c_K(t)
    =
    Z_K\Pi_{PP}^{\ell s}(m_K^2)
    \left[1+2G_\pi\Pi_{PP}^{\ell s}(m_K^2)\right]
    =0,
\label{eq:cbar_cancellation_kaon}
\end{equation}
where the final equality follows from the kaon pole condition,
Eq.~\eqref{eq:PP_pole_condition}.
The cancellation holds for unequal light and strange constituent masses
and explicitly shows why the contact diagram cannot be omitted.

The total form factors can be separated into the two flavor sectors,
\begin{equation}
    A_K(t)=A_\ell^K(t)+A_s^K(t),
    \qquad
    C_K(t)=C_\ell^K(t)+C_s^K(t).
\label{eq:flavor_decomposition_A_C}
\end{equation}
This decomposition is used below to identify the effect of explicit
$SU(3)_F$ breaking.

\subsection{Projectors and WTI-preserving implementation}
For on-shell kaons, $P\cdot\Delta=0$ and $P^2=m_K^2-t/4$.
Defining
\begin{equation}
    \mathcal M_{PP}=P_\mu P_\nu\mathcal M_K^{\mu\nu},
    \qquad
    \mathcal M_g=g_{\mu\nu}\mathcal M_K^{\mu\nu},
\end{equation}
the conserved form factors are obtained from
\begin{align}
    C_K(t)
    &=
    \frac{\mathcal M_{PP}-P^2\mathcal M_g}{P^2t},
\label{eq:C_projector}
\\
    A_K(t)
    &=
    \frac{1}{2P^2}
    \left[\mathcal M_g+\frac{3}{2}t\,C_K(t)\right].
\label{eq:A_projector}
\end{align}

This sequence of operations is crucial when applying proper-time regularization.
Specifically, we use the GWTI to reduce inverse-propagator and contact terms
at the integrand level, performing tensor contractions algebraically before
introducing the Feynman-parameter integral and proper-time cutoff. 
If the proper-time cutoff were applied prior to this integrand-level reduction,
the momentum-shift dependence inherent to nonrenormalizable contact interactions
would yield a spurious local surface term in the $C$-term channel.
Such a regulator artifact violates the soft-pion low-energy theorem, shifting
$C_\pi(0)$ away from $-1$ even in the chiral limit.
By enforcing the GWTI reduction at the unregularized level, this regulator
 artifact is identically eliminated, rigorously guaranteeing $C_\pi(0)=-1$
 without requiring any ad hoc subtractions or parameter tuning.

\subsection{Form factors and flavor decomposition}
\begin{figure}[tp]
    \centering
    \includegraphics[width=0.75\linewidth]{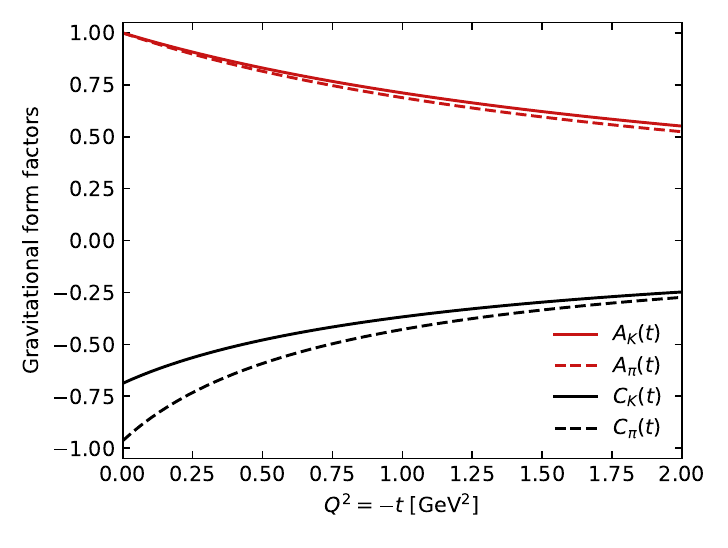}
    \caption{Kaon and pion gravitational form factors $A(t)$ and $C(t)$ as
    functions of $|t|$.}
    \label{fig:kaon_pion_gffs}
\end{figure}

Figure~\ref{fig:kaon_pion_gffs} compares the kaon and pion GFFs.
The momentum sum rule is satisfied for both mesons.
At zero momentum transfer, we obtain $C_\pi(0)=-0.965$ and $C_K(0)=-0.688$.
The kaon $D$-term is therefore substantially less negative than the pion value,
while the slopes of the two momentum form factors remain close.
The magnitude of $C_K(0)$ is compatible with recent continuum, dispersive, and
light-front analyses~\cite{Xu:2023ext,Cao:2025xhk,Sain:2026blfq}.

\begin{figure}[tp]
    \centering
    \includegraphics[width=0.75\linewidth]{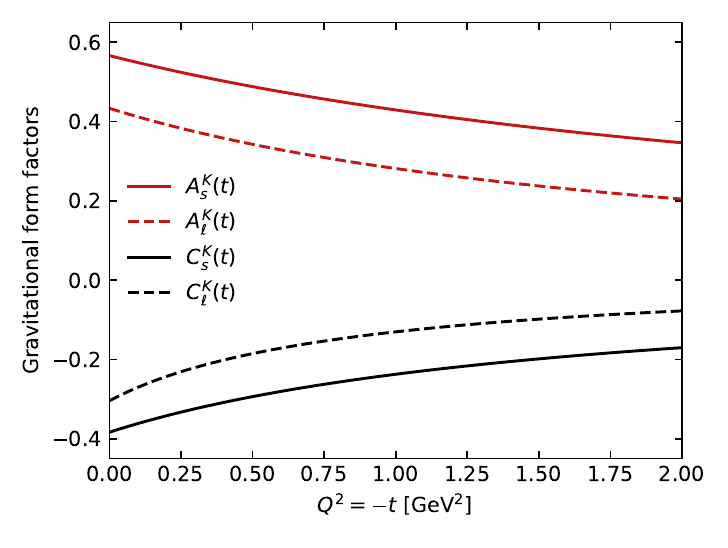}
    \caption{Light- and strange-quark contributions to the kaon form factors
    $A_K(t)$ and $C_K(t)$.}
    \label{fig:light_strange_gffs}
\end{figure}

The flavor-resolved form factors are displayed in Fig.~\ref{fig:light_strange_gffs}.
At $t=0$, we find $A_\ell^K(0)=0.434$ and $A_s^K(0)=0.566$, demonstrating that the
strange sector carries the larger fraction of the kaon momentum in the present model.
The corresponding $C$-term contributions are $C_\ell^K(0)=-0.304$ and $C_s^K(0)=-0.384$,
so the strange sector also gives the larger contribution to $|C_K(0)|$.
The different momentum-transfer dependences of the two contributions are a direct
manifestation of the unequal constituent masses.

\subsection{Pressure and shear-force distributions}
We obtain the Breit-frame pressure and shear-force distributions from $C_M(t)$ according to
\begin{align}
    p_M(r)
    &=
    -\frac{1}{6\pi^2}
    \int_0^\infty dQ\,
    \frac{Q^4}{2E_M(Q)}C_M(-Q^2)j_0(Qr),
\nonumber\\
    s_M(r)
    &=
    -\frac{3}{8\pi^2}
    \int_0^\infty dQ\,
    \frac{Q^4}{2E_M(Q)}C_M(-Q^2)j_2(Qr),
    \qquad
    E_M(Q)=\sqrt{m_M^2+Q^2/4}.
\label{eq:pressure_shear_def}
\end{align}
Although these static Breit-frame distributions are model- and frame-dependent, especially
for light mesons, they provide a useful common basis for comparing the pion and kaon.

The Fourier integrals require the form factor beyond the directly calculated spacelike grid.
We therefore use the fixed-normalization tripole continuation
\begin{equation}
    C_M^{\rm ext}(-Q^2)
    =
    \frac{C_M(0)}{\left(1+Q^2/\Lambda_M^2\right)^3},
\label{eq:C_tripole_extension}
\end{equation}
fitted to the calculated values over $0\leq Q^2\leq2~{\rm GeV}^2$.
The resulting scales are $\Lambda_\pi=1.81~{\rm GeV}$ and $\Lambda_K=2.11~{\rm GeV}$.
Coordinate-space quantities, particularly $p(0)$ and $s_{\max}$ for each meson $M$,
consequently retain a dependence on this ultraviolet continuation.

\begin{figure}[t]
    \centering
    \includegraphics[width=0.92\linewidth]{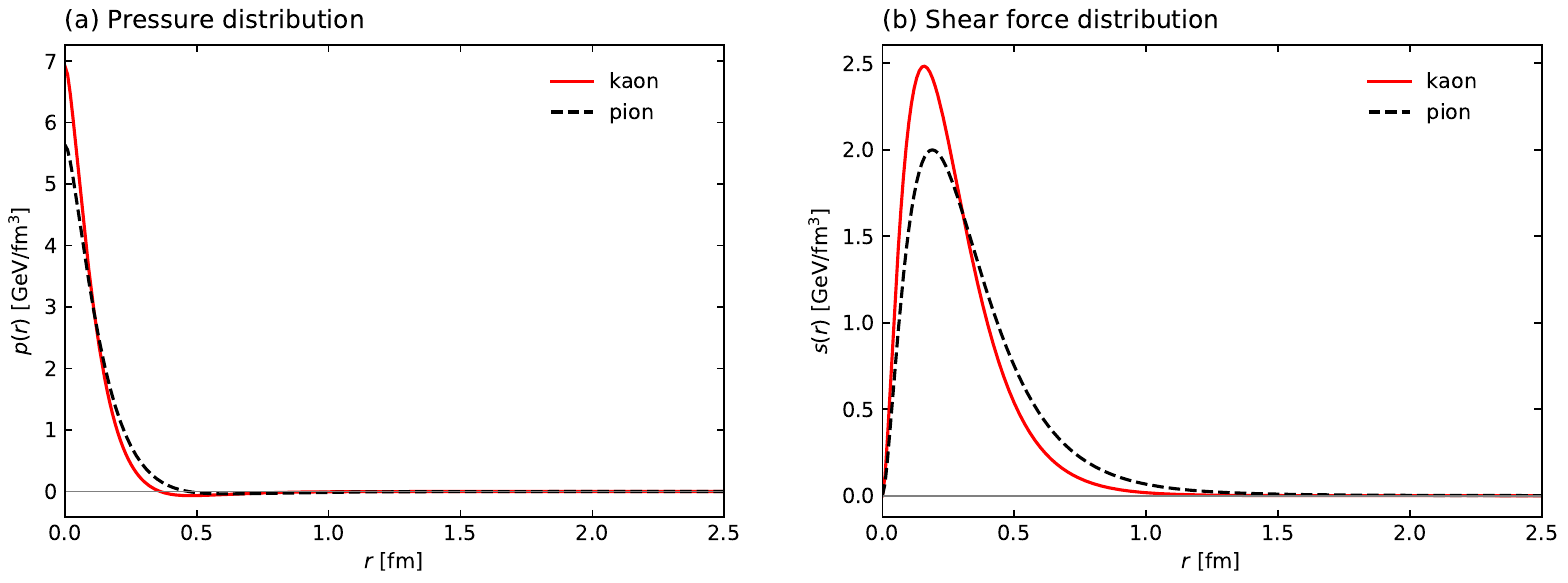}
    \caption{Breit-frame pressure $p(r)$ and shear-force $s(r)$ distributions
    of the pion and kaon.}
    \label{fig:pressure_shear}
\end{figure}

As shown in Fig.~\ref{fig:pressure_shear}, the kaon pressure and shear-force
distributions are more localized at smaller radii than the corresponding pion
distributions, with a larger central pressure and a larger peak shear force.
As summarized in Table~\ref{tab:mechanical_summary}, the first zero of the
pressure, $r_0$, moves from $0.47~\text{fm}$ for the pion to $0.36~\text{fm}$
for the kaon, and the position of the shear-force maximum shifts inward accordingly.
The peak value $s_{\rm max}$ increases by $25\%$, from $2.0~\text{GeV/fm}^3$ to
$2.5~\text{GeV/fm}^3$.
Within the tripole continuation of Eq.~\eqref{eq:C_tripole_extension}, these
changes characterize a more localized kaon stress distribution.
The local conservation relation $p(0)=2\int_0^\infty dr\,s(r)/r$ connects the
larger kaon central pressure, $p(0)=6.9~\text{GeV/fm}^3$, to the larger
shear-force distribution at small radii.
The $D$-term is related to the shear force through the mechanical sum rule
$C_M(0)=-\frac{16\pi m_M}{15}\int_0^\infty dr\,r^4s_M(r)$.
Because of the strong $r^4$ weighting in this spatial moment, the magnitude of
$C_M(0)$ is heavily sensitive to the tail of the stress distribution.
Although the kaon features a higher peak shear force $s_{\rm max}$ than the pion,
its mechanical distribution is concentrated at significantly smaller radial
distances.
This enhanced spatial compactness causes the $r^4$ factor to suppress the overall
integral, directly explaining why the kaon exhibits a noticeably less negative
$D$-term than the pion.

Finally, the von Laue condition for mechanical stability is satisfied within 
numerical accuracy for both mesons.
A qualitative trend toward a more compact kaon pressure profile is also found
in continuum and light-front studies~\cite{Xu:2023ext,Sain:2026blfq}; detailed
peak magnitudes are not directly comparable because the spatial representations
and ultraviolet completions differ.

\begin{table}[t]
\caption{Characteristic pressure and shear-force scales obtained with the
tripole continuation in Eq.~\eqref{eq:C_tripole_extension}; $r_0$ denotes
the first zero of $p(r)$.}
\label{tab:mechanical_summary}
\begin{ruledtabular}
\begin{tabular}{ccccc}
meson
&
$p(0)\,[{\rm GeV}/{\rm fm}^3]$
&
$r_0\,[{\rm fm}]$
&
$s_{\rm max}\,[{\rm GeV}/{\rm fm}^3]$
&
$r(s_{\rm max})\,[{\rm fm}]$
\\
\hline
$\pi$ & 5.6 & 0.47 & 2.0 & 0.19 \\
$K$   & 6.9 & 0.36 & 2.5 & 0.16
\end{tabular}
\end{ruledtabular}
\end{table}

\subsection{Mass radii}
In the Breit frame, where the energy transfer vanishes
($\Delta^0=0$ and $t=-\boldsymbol{\Delta}^2$), the spatial distributions of the
energy-momentum tensor are obtained via a three-dimensional Fourier transform,
\begin{equation}
    \langle T^{\mu\nu}\rangle_{\rm B}(\mathbf r)
    =
    \int\frac{d^3\boldsymbol{\Delta}}{2P^0(2\pi)^3}
    e^{-i\boldsymbol{\Delta}\cdot\mathbf r}
    \langle M(p')|T^{\mu\nu}(0)|M(p)\rangle
    \bigg|_{\Delta^0=0}.
\label{eq:Breit_density}
\end{equation}
The corresponding mean-square mass radius is defined as the second moment
of $T^{00}(\mathbf r)$, yielding
\begin{equation}
    \langle r_M^2\rangle_{\rm B}
    =
    6\left.\frac{dA_M(t)}{dt}\right|_{t=0}
    -
    \frac{3}{4m_M^2}\left[1+2C_M(0)\right].
\label{eq:kaon_Breit_mass_radius}
\end{equation}

Alternatively, a boost-invariant transverse density can be defined
in the Drell--Yan frame ($\Delta^+=0$) through a two-dimensional Fourier transform,
\begin{equation}
    \langle T^{++}\rangle_{\rm LF}(\mathbf r_\perp)
    =
    \int\frac{d^2\boldsymbol{\Delta}_\perp}{2P^+(2\pi)^2}
    e^{-i\boldsymbol{\Delta}_\perp\cdot\mathbf r_\perp}
    \langle M(p')|T^{++}(0)|M(p)\rangle
    \bigg|_{\Delta^+=0}.
\label{eq:LF_density}
\end{equation}
Its mean-square transverse radius is defined as the second moment of $T^{++}(\mathbf r_\perp)$,
which depends solely on the slope of $A_M(t)$~\cite{Burkardt:2000za,Cao:2024lc}:
\begin{equation}
    \langle r_{M,\perp}^2\rangle_{\rm LF}
    =
    4\left.\frac{dA_M(t)}{dt}\right|_{t=0}.
\label{eq:kaon_LF_mass_radius}
\end{equation}

\begin{table}[t]
\caption{Breit-frame and transverse light-front mass radii.}
\label{tab:mass_radii}
\begin{ruledtabular}
\begin{tabular}{ccc}
meson
&
$\sqrt{\langle r^2\rangle_{\rm B}}\,[{\rm fm}]$
&
$\sqrt{\langle r_\perp^2\rangle_{\rm LF}}\,[{\rm fm}]$
\\
\hline
$\pi$ & 1.24  & 0.266 \\
$K$   & 0.381 & 0.252
\end{tabular}
\end{ruledtabular}
\end{table}

The radii are summarized in Table~\ref{tab:mass_radii}.
The unusually large Breit-frame pion radius originates from the term proportional
to $[1+2C_\pi(0)]/m_\pi^2$ in Eq.~\eqref{eq:kaon_Breit_mass_radius};
it is therefore highly sensitive to the small pion mass.
By contrast, the transverse light-front radius depends only on the slope
of $A_\pi(t)$ and remains close to the kaon value, providing a robust,
frame-independent measure of the intrinsic distribution of energy and
momentum inside Goldstone bosons.

\section{Summary and discussion}\label{sec:summary}
We have presented a symmetry-preserving calculation of the kaon gravitational
form factors in a proper-time-regularized three-flavor NJL model.
Three key ingredients implement the physical constraints.
First, the quark--graviton vertex is dressed consistently with the four-fermion
interaction to strictly satisfy the GWTI; together with canonical
Bethe--Salpeter normalization, this ensures $A_K(0)=1$.
Second, the interaction-induced bicycle diagram is included together with
the light- and strange-quark triangle diagrams, and the on-shell kaon pole
condition cancels the nonconserved form factor $\bar c_K(t)$.
Third, for the $C$-term form factor, executing the GWTI algebraic reduction at
the integrand level \emph{prior} to imposing the proper-time cutoff removes
a local regulator artifact, thereby rigorously preserving the soft-pion
low-energy theorem ($C_\pi(0)=-1$) in the chiral limit without requiring
ad hoc subtractions.

The strange constituent carries a dominant share of both the kaon momentum
and its mechanical response.
Consequently, the total kaon $D$-term is noticeably less negative than the
pion value, despite the slopes of their momentum form factors $A(t)$
remaining quite similar.
We therefore conclude that the primary kaon--pion difference in this framework
is not a uniform spatial rescaling, but rather a fundamental reshaping of the
internal stress response driven by unequal constituent masses.

This distinction is also visible in coordinate space.
The kaon pressure and shear-force profiles are more compact and attain larger
central or peak values than the pion profiles, while both satisfy the von Laue
stability condition. 
The comparison between Breit-frame and transverse light-front mass radii
sharply highlights different facets of meson structure:
while the Breit-frame radius offers an intuitive three-dimensional static picture,
it is highly sensitive to the small physical pion mass through the term
$[1+2C_\pi(0)]/m_\pi^2$.
In contrast, the boost-invariant light-front radius probes the geometric
transverse size---depending solely on the slope of $A(t)$---and yields
similar sizes for both mesons.
This contrast underscores the necessity of specifying the reference frame
and density operator when assigning a spatial interpretation to the GFFs of
light Nambu--Goldstone bosons.

It is instructive to compare these findings with recent continuum Schwinger-function
and basis light-front quantization (BLFQ) studies~\cite{Xu:2023ext,Sain:2026blfq}.
Both approaches likewise find a less negative kaon $D$-term and a more compact stress
distribution than in the pion case.
In the continuum calculation, dynamical chiral symmetry breaking and the heavier
strange-quark mass lead to a more localized kaon stress distribution and enhanced
central pressure~\cite{Xu:2023ext}.
By contrast, BLFQ employs a trivial light-front vacuum and encodes chiral symmetry
breaking phenomenologically through constituent masses and effective potentials.
Nevertheless, its flavor-asymmetric light-front wave functions also assign the
strange sector a larger momentum fraction and a larger share of the mechanical
response~\cite{Sain:2026blfq}.
The persistence of this qualitative ordering across distinct frameworks confirms
that the enhanced compactness and modified mechanical response of the kaon are
robust consequences of explicit $SU(3)_F$ symmetry breaking.

Finally, we note the intrinsic limitations of the present framework.
As a leading-order low-energy quark-model evaluation, it lacks explicit
gluon degrees of freedom, omits the Kobayashi--Maskawa--'t Hooft determinant
interaction, and neglects meson-loop corrections.
The flavor-separated form factors should therefore be interpreted at the
intrinsic model scale rather than at a generic perturbative-QCD scale,
and the coordinate-space stress profiles additionally depend on the choice
of ultraviolet extrapolation scheme.
Natural extensions include the Kobayashi--Maskawa--'t Hooft interaction,
subleading $1/N_c$ corrections, and a matching and evolution analysis that
permits direct comparison with lattice QCD and phenomenological extractions.
These developments would help clarify which features identified here are
consequences of symmetry and which depend on the contact-interaction dynamics.

\section*{Acknowledgments}

The work of Z.L. was supported by the THERS Make New Standards Program for the
Next Generation Researchers at Nagoya University (No.~2562409026) and by
JSPS KAKENHI Grant-in-Aid for JSPS Fellows (JSPS DC2, Grant
No.~JP26KJ1338).

\bibliographystyle{apsrev4-2}
\bibliography{reference}

\end{document}